\documentclass[twocolumn,showpacs,preprintnumbers,amsmath,amssymb,10pt]{revtex4}

\usepackage{psfig}
\usepackage{graphicx}
\usepackage{dcolumn}
\usepackage{bm}
\newcommand{\s}{\ensuremath{\psi(t,r)}}
\newcommand{\n}{\ensuremath{\nu(t,r)}}
\newcommand{\T}{\ensuremath{\theta}}
\newcommand{\pt}{\ensuremath{p_\theta}}
\newcommand{\pr}{\ensuremath{p_r}}
\newcommand{\e}{equation} 
\newcommand{\M}{\ensuremath{{\cal M}}}
\newcommand{\prz}{\ensuremath{p_{r_0}}}

\newcommand{\ptz}{\ensuremath{p_{\theta_0}}}

\newcommand{\X}{\ensuremath{{\cal X}}}

\begin{document}

\preprint{}

\title{Role of initial data in spherical collapse}

\author{ Pankaj S Joshi}
\email{psj@tifr.res.in}
\author{Rituparno Goswami}
\email{goswami@tifr.res.in}
\affiliation{ Department of Astronomy and Astrophysics\\ Tata 
Institute of Fundamental Research\\ Homi Bhabha Road,
Mumbai 400 005, India}

\begin{abstract}

We bring out here the role of initial data in causing the black hole
and naked singularity phases as the final end state of a continual
gravitational collapse. The collapse of a type I general matter field 
is considered, and it is shown that given the distribution of density and
pressure profiles at the initial surface from which the collapse
evolves, there is a freedom in choosing rest of the free functions
and dynamical collapse evolutions from this initial data,
such as the velocities of the collapsing shells, so that the end state
could be either a black hole or a naked singularity depending on this 
choice. Our results give some physical insights into these collapse 
outcomes as to how these phases come about.

\end{abstract}

\pacs{04.20.Dw, 04.70.-s, 04.70.Bw}

\maketitle

An intriguing finding that has emerged from recent investigations 
on gravitational collapse scenarios in general relativity is that the
final end state of such a collapse could be either a black hole (BH),
or a naked singularity (NS). The theoretical and observational properties
of these objects could be quite different from each other and so it is of 
immense interest to get an insight into how each of these phases come 
about as end states of a dynamically developing collapse ruled 
by gravity.  
When a star is sufficiently massive, and starts collapsing 
gravitationally on exhausting its nuclear fuel, it would not settle to 
a stable configuration such as a neutron star. What happens in 
such a case is an endless gravitational collapse ensues, where 
the sole governing force is gravity. Then either an event horizon of 
gravity forms well in advance to create a black hole, or in the case 
otherwise, the extreme
density regions may fail to be covered by the horizon, and a visible
naked singularity may develop (see e.g. [1] for some recent reviews). 
Most of the models studied so far are spherical, and the matter
cloud collapses under reasonable physical conditions such as an
energy condition, and regularity of the initial data from which
the collapse develops.
The possible physical consequences of the later scenario has
drawn some attention recently [2].

The major unresolved problem, however, is to understand what 
causes these BH/NS phases as end states of a continual collapse, 
and how these
come about. In other words, we need to know why and when a naked 
singularity results as collapse end state as opposed to a black hole.
We study here collapse for general {\it Type I} matter fields, and our
purpose is to show explicitly that there exist classes of
solutions of Einstein equations which take the collapse evolution
to either of the BH/NS phases, 
subject to usual regularity and energy conditions, and depending on the 
choice of the free functions available.
While we discuss spherically 
symmetric collapse, the form of matter we consider is rather 
generic which is any type I general matter field subject to 
an energy condition [3]. What is seen here is given the initial matter 
distribution from which the collapse develops, there is a freedom to 
choose the velocity function (and its derivatives) for the collapsing 
shells, so that the collapse could end in either a black hole or naked 
singularity, depending on this choice. We thus get here some insight 
into what causes these phases, in terms of physical agencies such 
as initial matter and velocity distributions, and the allowed 
possible dynamical collapse evolutions.

The method we follow here is outlined below.
In the case of a black hole developing as collapse end state,
the spacetime singularity is necessarily hidden behind the event
horizon of gravity, whereas in the case of a naked singularity
developing, there are families of future directed non-spacelike 
trajectories coming
out from the singularity, which can in principle communicate
information to faraway observers in the spacetime. The existence
of such families confirms the NS formation, as opposed to a BH
end state. We shall study the
singularity curve produced by the collapsing matter, the tangent to
which at the central singularity at $r=0$ is related to the radially
outgoing null geodesics from the singularity if there are any,
as we show here.
By determining the nature of the singularity
curve and its relation to the initial data and the classes of 
collapse evolutions, we are 
able to deduce
whether the trapped surface formation in collapse takes place
before or after the singularity. It is this causal 
structure of the trapped region that determines the possible emergence or 
otherwise of 
non-spacelike paths from the singularity. This 
settles the final outcome of collapse in terms of either a BH or NS. 
We point out that 
several familiar equations of state for which extensive collapse studies 
have been made so far, such as dust, matter with only tangential or radial
pressures, and others form special cases of the 
considerations here.

The spacetime geometry within the spherically symmetric collapsing 
cloud can be described by the general metric in the comoving   
coordinates $(t,r,\theta,\phi)$ as given by,
\begin{equation}
ds^2=-e^{2\n}dt^2+e^{2\s}dr^2+R^2(t,r)d\Omega^2
\label{eq:metric}
\end{equation}
where $d\Omega^2$ is the line element on a two-sphere. 
The energy-momentum tensor is that for any matter field 
which is type I (this 
is a broad class which includes most of the physically reasonable matter
fields, including dust, perfect fluids, massless scalar fields and such
others, see e.g. [3]), and is given in a diagonal form,  
$T^t_t=-\rho, T^r_r=p_r, T^\T_\T=T^\phi_\phi=p_\T$.
The quantities 
$\rho$, $p_r$ and $p_\T$ are the energy density, radial 
and tangential pressures respectively. We take the matter field 
to satisfy the {\it weak energy condition}, i.e. the energy density 
measured by any local observer be non-negative, and so for any 
timelike vector $V^i$, we must have $T_{ik}V^iV^k\ge0$,
which amounts to $\rho\ge0;\; \rho+p_r\ge0;\; \rho+p_\T\ge0$.

Now for the metric (\ref{eq:metric}) the Einstein equations 
take the form, in the units $(8\pi G=c=1)$

\begin{eqnarray}
\rho=\frac{F'}{R^2R'}; && p_r=-\frac{\dot{F}}{R^2\dot{R}}
\label{eq:ein1}
\end{eqnarray}
\begin{equation}
\nu'=\frac{2(\pt-p_r)}{\rho+p_r}\frac{R'}{R}-\frac{p_r'}{\rho+p_r}
\label{eq:ein2}
\end{equation}
\begin{equation}
-2\dot{R}'+R'\frac{\dot{G}}{G}+\dot{R}\frac{H'}{H}=0
\label{eq:ein3}
\end{equation}
\begin{equation}
G-H=1-\frac{F}{R}
\label{eq:ein4}
\end{equation}
where we have defined $G(t,r)=e^{-2\psi}(R')^2$ and 
$H(t,r)=e^{-2\nu}(\dot{R})^2$.

The arbitrary function $F=F(t,r)$ has an interpretation of
the mass function for the cloud, giving the total mass in a 
shell of comoving radius $r$. Energy conditions imply $F\ge0$. 
In order to preserve the regularity at the initial epoch, we have  
$F(t_i,0)=0$, that is the mass function should vanish at the center 
of the cloud. As seen from equation (2), there is a density singularity
in the spacetime at $R=0$, and at $R'=0$. However, the later one 
is due to shell-crossings and can be possibly removed from the 
spacetime [4], so we shall consider here only the shell-focusing 
singularity at $R=0$, which is a physical singularity where all
matter shells collapse to a zero physical radius.
We can use the scaling freedom available for the radial co-ordinate 
$r$ to write $R=r$ at the initial epoch $t=t_i$. Introducing the
function $v(t,r)$ by the relation,
\begin{equation}
R(t,r)=rv(t,r)
\label{eq:R}
\end{equation}
we have $v(t_i,r)=1; v(t_s(r),r)=0$  and for collapse, $\dot{v}<0$.
The time $t=t_s(r)$ corresponds to the shell-focusing singularity 
$R=0$, where all the matter shells collapse to a vanishing physical
radius.

From the point of view of dynamic evolution of the initial data
prescribed at the initial epoch $t=t_i$, there are six arbitrary functions 
of the comoving shell-radius $r$ as given by,
$\nu(t_i,r)=\nu_0(r), \psi(t_i,r)=\psi_0(r), R(t_i,r)=r 
\rho(t_i,r)=\rho_0(r), p_r(t_i,r)=p_{r_0}(r)
\pt(t_i,r)=p_{\T_0}$.
We note that not all the initial data above are mutually independent, 
because from equation (\ref{eq:ein2}) we get,
\begin{equation}
\nu_0(r)=\int_0^r\left(\frac{2(p_{\T_0}-p_{r_0})}{r(\rho_0+p_{r_0})}
-\frac{p_{r_0}'}{\rho_0+p_{r_0}}\right)dr
\label{eq:nu0}
\end{equation}
Let us now assume that the initial pressures have physically 
reasonable behavior at the center $(r=0)$ in that the pressure gradients 
vanish, i.e. $\prz'(0)=\ptz'(0)=0$, and also that the difference between 
radial and tangential pressures vanishes at the center, 
i.e. $\prz(0)-\ptz(0)=0$, to ensure the regularity of the initial data 
at the center of the cloud.
Then from \e(\ref{eq:nu0}), it is evident that $\nu_0(r)$ has the form,
\begin{equation}
\nu_0(r)=r^2g(r)
\label{eq:nu0form}
\end{equation} 
where, $g(r)$ is at least a $C^1$ function of $r$ for $r=0$, and at least 
a $C^2$ function for $r>0$.

Hence we see that there are five total field equations with seven 
unknowns, $\rho$, $p_r$, $\pt$, $\psi$, $\nu$, $R$, and $F$, giving us 
the freedom of choice of two free functions. Their selection, subject to 
the weak energy condition and the given initial 
data for collapse at the initial surface, 
determines the matter distribution and metric of the space-time, and 
thus leads to a particular time evolution of the initial data.

Consider now a general mass function $F(t,r)$ for the collapsing
cloud, which can be written as
\begin{equation}
F(t,r)=r^3\M (r,v)
\label{eq:mass}
\end{equation}
Here $\M$ is a regular and suitably differentiable function and 
from equations below, the regularity and finiteness of the density 
profile at the initial epoch $t=t_i$ requires that $F$ goes as $r^3$ 
close to the center. 
The equations (\ref{eq:ein1}) and (6) give, 
\begin{eqnarray}
\rho=\frac{3\M+r\left[\M_{,r}+\M_{,v}v'\right]}{v^2(v+rv')};
&&\pr=-\frac{\M_{,v}}{v^2}
\label{eq:rho} 
\end{eqnarray}
Thus regular density distribution at the initial epoch is given by,
$\rho_0(r)=3\M(r,1)+r\M(r,1)_{,r}$. 
It is evident that, in general, as 
$v\rightarrow 0$, $\rho\rightarrow\infty$ and $\pr\rightarrow\infty$. 
That is, both the 
density and radial pressure blow up at the singularity.

What we show below is, given any regular initial density
and pressure profiles for the matter cloud from which the collapse 
develops, there always exist velocity profiles for collapsing matter 
shells, and classes of dynamical evolutions as determined by the 
Einstein equations, so that the end state of the collapse would
be either a naked singularity or a black hole, depending  
on the choice made. In other words, given the matter initial data at the 
initial surface $t=t_i$, these evolutions take the collapse to end up 
either as a black hole or the naked singularity, depending on the choice 
of the class.

To see this, we provide now classes of solutions to
Einstein equations to this effect. Consider the class of velocity profiles 
as determined (using equations (7) and (8)) by the general function,
\begin{equation}
\n=A(t,R)
\label{eq:nu}
\end{equation}
where $A(t,R)$ is any arbitrary, suitably differentiable 
function of $t$ and $R$, with the initial 
constraint $A(t_i,R)=\nu_0(r)$.
Using \e (\ref{eq:nu}) in \e (\ref{eq:ein3}), we have,
\begin{equation}
G(t,r)=b(r)e^{2(A-\int A_{,t}dt)}
\label{eq:G}
\end{equation}
Here $b(r)$ is another arbitrary function of $r$. A comparison
with dust models interprets $b(r)$ as the velocity function for
the shells. (It is important to note here that we could have 
also chosen $\nu(t,r)$ of the form $\nu'=a(r,v)_{,v}R'$, where
$a(r,v)$ is again an arbitrary and suitably differenciable
function, which too would allow us to integrate equation (\ref{eq:ein3})). 
From equation (\ref{eq:nu0form}) let us generalize the form 
of $A(t,R)$ as, $A(t,R)=r^2g_1(r,v)$, where $g_1(r,v)$ is a suitably 
differentiable function and $g_1(r,1)=g(r)$.
Then we similarly have $A-\int A_{,t}dt=r^2g_2(r,v)$
and at the initial epoch $g_2(r,1)=g(r)$.
Using \e(\ref{eq:nu}) in \e(\ref{eq:ein2}), we get,
\begin{equation}
2\pt=RA_{,R}(\rho+\pr)+2\pr+\frac{Rp_r'}{R'}
\label{eq:ptheta}
\end{equation}
In general, both the density and radial pressure blow up at 
the singularity, so the above equation implies that the tangential 
pressure also blows up at the singularity.
Writing
\begin{equation}
b(r)=1+r^2b_0(r)
\label{eq:veldist}
\end{equation}
and using \e s(\ref{eq:mass}),(\ref{eq:nu}) and (\ref{eq:G}) 
in \e(\ref{eq:ein4}), we get,
\begin{equation}
\sqrt{R}\dot{R}=-e^{r^2g_1(r,v)}\sqrt{(1+r^2b_0)Re^{r^2g_2(r,v)}-R+r^3\M}
\label{eq:collapse}
\end{equation}
Since we are considering collapse, we have $\dot{R}<0$.
Defining a function $h(r,v)$ as,
\begin{equation}
h(r,v)=\frac{e^{2r^2g_2(r,v)}-1}{r^2}=2g_2(r,v)+{\cal O}(r^2v^2)
\label{eq:h}
\end{equation}
and using \e(\ref{eq:h}) in \e(\ref{eq:collapse}), we get 
after simplification,
\begin{equation}
\sqrt{v}\dot{v}=-\sqrt{e^{2r^2(g_1+g_2)}vb_0+e^{2r^2g_1}\left(vh(r,v)
+\M(r,v)\right)}
\label{eq:collapse1}
\end{equation}
Integrating the above equation we have,
\begin{equation}
t(v,r)=\int_v^1\frac{\sqrt{v}dv}{\sqrt{e^{2r^2(g_1+g_2)}vb_0+e^{2r^2g_1}
\left(vh+\M\right)}}
\label{eq:scurve1}
\end{equation}
Note that the variable $r$ is treated as a constant in the above equation. 
Expanding $t(v,r)$ around the center, we get,
\begin{equation} 
t(v,r)=t(v,0)+r\X(v)+{\cal O}(r^2)
\label{eq:scurve2}
\end{equation}
where the function $\X(v)$ is given by,
\begin{equation}
\X(v)=-\frac{1}{2}\int_v^1dv\frac{\sqrt{v}(b_1v+vh_1(v)+\M_1(v))}
{\left(b_{00}v+vh_0(v)+\M_0(v)\right)^{\frac{3}{2}}}
\label{eq:tangent1}
\end{equation}
with, $b_{00}=b_0(0), M_0(v)=\M(0,v), 
h_0=h(0,v), b_1=b_0'(0), M_1(v)=\M_{,r}(0,v), h_1=h_{,r}(0,v)$.

From the above, the time when the central singularity develops is 
given by
\begin{equation}
t_{s_0}=\int_0^1\frac{\sqrt{v}dv}{\sqrt{b_{00}v+vh_0(v)+\M_0(v)}}
\label{eq:scurve3}
\end{equation}
The time for other shells to reach the singularity  
can be given by the expansion,
\begin{equation}
t_s(r)=t_{s_0}+r\X(0)+{\cal O}(r^2)
\label{eq:scurve4}
\end{equation}
It is now clear that the value of $\X(0)$ depends 
on the functions $b_0,\M$ and $h$, which in turn depend on 
the initial data at $t=t_i$ and the dynamical variable $v$. 
Thus, a given set of initial matter distribution 
and the dynamical profiles including the velocity of shells
completely determine the tangent at the center
to the singularity curve.
Further, from equations (17-19), we get,
\begin{equation}
\sqrt{v}v'=\X(v)\sqrt{b_{00}v+vh_0(v)+\M_0(v)}+{\cal O}(r^2)
\label{eq:scurve5}
\end{equation}

It is now possible to see how the final fate of collapse 
is determined in terms of either a black hole or 
a naked singularity. This is determined by the causal behavior of
the apparent horizon, which is the boundary of 
trapped surfaces forming due to collapse, and is given by $R=F$. 
If the neighborhood 
of the center gets trapped prior to the epoch of singularity, then it 
is covered and a black hole results, otherwise it could be naked when  
non-spacelike future directed trajectories escape from it.
The key point is to determine if there are any future-directed
non-spacelike paths emerging from the singularity. To see
this, and to examine the nature of the central singularity at $R=0$, 
$r=0$, consider the equation for outgoing radial null geodesics,
\begin{equation} 
\frac{dt}{dr}=e^{\psi-\nu}
\end{equation}
We want to examine if there would be any families of future
directed null geodesics
coming out of the singularity, thus causing a NS phase.

The singularity occurs at $v(t_s(r),r)=0$, i.e. $R(t_s(r),r)=0$. 
Therefore, if there are any future directed null geodesics, 
terminating in the past at the singularity, we must have 
$R\rightarrow0$ as $t\rightarrow t_s$ along these curves. Now writing 
the null geodesics equation in terms of the variables $(u=r^\alpha,R)$, 
choosing $\alpha=\frac{5}{3}$, and using \e (\ref{eq:ein4}) 
we get,
\begin{equation}
\frac{dR}{du}=\frac{3}{5}\left(\frac{R}{u}+\frac{\sqrt{v}v'}
{\sqrt{\frac{R}{u}}}\right)\left(\frac{1-\frac{F}{R}}
{\sqrt{G}[\sqrt{G}+\sqrt{H}]}\right)
\label{eq:null3}
\end{equation}

If the null geodesics terminate at the singularity in the 
past with a definite tangent, then at the singularity we have 
$\frac{dR}{du}>0$, in the $(u,R)$ plane with a finite value. 
Hence all points $r>0$ on the singularity curve are covered 
since $F/R \rightarrow\infty$ with $\frac{dR}{du}\rightarrow-\infty$. 
The central singularity could however be naked. 
Define the tangent to the outgoing null geodesic from the singularity as,
\begin{equation}
x_0=\lim_{t\rightarrow t_s}\lim_{r\rightarrow 0} \frac{R}{u}
=\left.\frac{dR}{du}\right|_{t\rightarrow t_s;r\rightarrow 0}
\end{equation}
Using \e (\ref{eq:null3}) and (\ref{eq:scurve5}) we then get,
\begin{equation}
x_0^{3/2}= (3/2)\sqrt{\M_0(0)}\X(0)
\label{eq:null4}
\end{equation}
and the equation
of null geodesic emerging from the singularity is $R=x_0u$, or
in $(t,r)$ coordinates it is given by 
\begin{equation} 
t-t_s(0)=x_0r^{5/3}
\label{eq:null5}
\end{equation}
It follows now that if $\X(0)>0$, then $x_0>0$, and we 
get radially outgoing null geodesics coming out from the singularity, 
giving rise to a naked central singularity.
However, if $\X(0)<0$ we have a black hole solution, as there will
be no such trajectories coming out. If $\X(0)=0$ then we will have 
to take into account the next higher order non-zero term in the 
singularity curve equation, and a similar analysis has to be carried out by 
choosing a different value of $\alpha$.

We note that in the above the functions $h$ and ${\cal M}$ are
expanded with respect to $r$ around $r=0$ and the first-order terms
are considered. At times, however, these are assumed to be
expandable with respect to $r^{2}$, and it is 
argued that such smooth functions would be physically more relevant.
Such an assumption comes from the analyticity with respect to the  
local Minkowskian coordinates (see e.g. [5]), and it is really
the freedom of definition mathematically. We may remark that the 
formalism as discussed here would work for such smooth functions 
also, which is a special case of above.

We thus see how the initial data 
in terms of the free functions available determine the 
BH/NS phases as collapse end states, because $\X(0)$ is determined
by these initial and dynamical profiles as given by 
\e(\ref{eq:tangent1}). It is clear, therefore,
that given any regular density and pressure profiles for
the matter cloud from which the collapse develops, we can always 
choose velocity profiles so that the end state of the collapse would
be either a naked singularity or a black hole, and vice-versa.
It is interesting to note here that physical agencies such as the
spacetime shear within a dynamically collapsing cloud could naturally
give rise to such phases in gravitational collapse [6].
In other words, such physical factors can naturally delay the
formation of apparent horizon and the trapped surfaces.

We believe numerical work on collapse models may provide further 
insights into these interesting dynamical phenomena, and especially 
when collapse is non-spherical, which 
remains a major problem to be considered. 
Much numerical work has been done in recent years on spherical scalar 
field collapse [7], and also on some perfect fluid models [8]. 
While we have worked out explicitly here the
emergence of null geodesics from the singularity, thus showing it to be
naked (or otherwise) in an analytic manner, the numerical
simulations generally discuss the formation or otherwise of
trapped surfaces and apparent horizon, and these may also break down 
closer to the epoch of actual singularity formation. In that case
this may not allow for actual detection of BH/NS end 
states, whereas important insights on critical phenomena 
and dispersal have already been gained through numerical methods. 
Probably, a detailed numerical investigation of the null geodesics 
in collapse may provide further insights here.

As stated above, we work here with {\it type I} matter fields, 
which is a rather general form of matter which includes
practically all known physically reasonable fields such as dust,
prefect fluids, massless scalar fields and so on. However, it is
important to note that suitable care must be taken in interpreting our 
results. While we have shown that the initial data and dynamical 
evolutions chosen do determine the BH/NS end states for collapse, the 
point is, actually, all these dynamical variables are not explicitly 
determined by the initial data given at the initial epoch (note 
that $v$ plays the role of a time coordinate here). Hence these functions 
are determined only as a result of time development of the system 
from the initial data if we have the relation between the density and 
pressures, that is an `equation of state'. In principle, it is possible 
to choose these functions freely, only subject to an energy condition,
and after that one may calculate the 
energy density, and the radial and tangential pressures for the matter. 
However, in that case the resultant `equation of state' could be 
quite strange in general. Of course, presently we have practically 
very little idea on what kind of an equation of
state should the matter follow, especially closer to the collapse 
end states, where we are already dealing with ultra-high energies
and pressures. Hence if we allow for the possibility that we 
could freely choose the property of the matter fields as above, or 
the equation of state, then our analysis is certainly valid. In such 
a case, however, the chosen `equation of state' will be in general such  
that the pressures may explicitly depend not only on the energy 
density, but also on the time coordinate.

All the same, it is worth pointing 
out that the analysis as given above in fact does include several 
well-known classes of collapse models and equation of states as we show 
below. It has already been stated that we have a choice of two free 
functions available here, among which we are choosing 
only one, namely, the form of metric function $\n$. Therefore, in addition 
to the Einstein equations, if any equation of state of the form 
$p_r=f(\rho)$ (or alternatively, $p_\theta=g(\rho)$)
is given, then it is clear from equation (\ref{eq:ein1}) that 
there would be a constraint on the otherwise arbitrary function $\M$,
specifying the required class, if the solution of the constraint equation 
exists. Then the value of $p_\theta$ (or $p_r$) in terms of 
$\rho(t,r)$ is determined from equation (\ref{eq:ein2}), and the 
analysis for the BH/NS phases as above still goes through. 
For example, for dust collapse models we have
\begin{equation} 
p_r=p_\theta=0
\end{equation} 
and the constraint equation gives,
\begin{equation} 
\M(r,v)=\M(r)
\end{equation} 
Next, for the collapse with a constant (or zero)
radial pressure, but with a variable tangential pressures allowed,
we have, from the constraint equation,
\begin{equation}
\M(r,v)=f(r)-kv^3
\end{equation}
where, $k$ is the value of the constant radial pressure. The tangential
pressure will then be given by ,
\begin{equation}
p_\theta=k+\frac{1}{2}A_{,R}R(\rho+k)
\end{equation}

Along with these two well-known models, the above analysis works for
any other models also in which one of the pressures is specified by an 
equation of state, and which permits a solution to the constraint equation 
on $\M$. Hence it follows that the considerations above provide
an interesting framework for the study of dynamical collapse, which
is one of the most important open problems in gravity physics today.

We note here that in the consideration above, we have dealt
with the {\it local} visibility or otherwise of the singularity,
which is the question of {\it strong cosmic censorship}. Such a locally 
naked singularity may or may not be {\it globally} visible, which could 
communicate to faraway observers in the spacetime (thus violating 
the {\it weak cosmic censorship}). This is actually decided
by the global behaviour (for large values of $r$) of the concerned 
functions such as the mass function and others. Typically it is 
observed (see e.g. [1]) that once the singularity is locally visible, 
it is possible
to choose the behaviour of functions away from the center in such a 
manner that it would be globally naked as well. A choice otherwise
could make all the non-spacelike trajectories coming
out of the singularity to fall back into the apparent horizon,
thus giving only a locally naked singularity.

The basic result we proved is, given a regular initial 
data in terms of the density and pressure profiles at the initial 
epoch from which the collapse develops, there are sets of dynamical  
evolutions, i.e. classes of solutions to the Einstein equations, which
can evolve the given initial data to produce either a black hole or a
naked singularity as the collapse end state. The exact outcome in 
terms of BH/NS end states depends on the choice of rest of the free 
functions available, such as the velocities of the collapsing shells.
In other words, the initial data space can be divided into distinct 
subspaces, those that evolve into black holes, and others into a 
naked singularity. In a sense this
brings out the stability of these collapse outcomes with respect to
perturbations in the initial data, and the choice of matter that 
collapses. This would be useful information on the important issue of 
genericity for collapse outcomes. Of course stability theory
in general relativity is a complex issue and what is needed is a
study of non-spherical collapse.
While some insights have been provided earlier by specific
examples, the generalization as given here may be useful 
in view of the significance of this problem in black hole physics, and 
for relativistic astrophysics in general. 
Further details shall be presented elsewhere.


\begin{references}


\bibitem{1}
For recent reviews, see, e.g. A. Krolak, Prog. Theor. Phys. Suppl. 
{\bf 136}, 45 (1999); R. Penrose, in {\it Black holes and relativistic stars},
ed. R. M. Wald (University of Chicago Press, 1998); P. S. Joshi and 
I. H. Dwivedi, Class.Quantum Grav. {\bf 16}, 41 (1999); P. S. Joshi, 
Pramana {\bf 55}, 529 (2000); M. Celerier and P. Szekeres, gr-qc/0203094;
R. Giambo', F. Giannoni, G. Magli, P. Piccione, gr-qc/0204030.

\bibitem{2}
T. Harada, H. Iguchi, and K. Nakao, Prog.Theor.Phys. {\bf 107} (2002) 449-524;
P. S. Joshi, N. Dadhich, and R. Maartens, Mod. Phys. Lett. 
{\bf A15}, 991 (2000); C. Vaz and L. Witten, Phys.Lett. {\bf B442} (1998) 90-96.

\bibitem{3}
S. W. Hawking and G. F. R. Ellis, {\it The large scale structure of
spacetime}, Cambridge University Press, Cambridge.

\bibitem{4}
C. J. S. C. Clarke, {\it Analysis of spacetime singularities},
Cambridge University Press, Cambridge.


\bibitem{5} D. Christodoulou, Commun. Math. Phys. {\bf 93}, 171 (1984). 


\bibitem{6}
E. Brinis, S. Jhingan and G. Magli, Class. Quantum Grav. {\bf 17} 
(2000) p. 4481; P. S. Joshi, N. Dadhich and R. Maartens, Phys. Rev. {\bf D65}
101501, 2002. 

\bibitem{7} See e.g. C. Gundlach, Living Rev. Relat., {\bf 2}, 4 (1999), 
and references there in for a review.

\bibitem{8}
A. Ori and T. Piran, Phys. Rev. Lett., 59, p.2137 (1987); 
Phys. Rev. D42, p.1068 (1990);
T. Foglizzo and R. Henriksen, Phys. Rev. D48, p.4645 (1993);
T. Harada, Phys. Rev. D58, p.104015 (1998); T. Harada and H. Maeda,
Phys. Rev. D63, p.084022 (2001).



\end{references}
\end{document}